\documentclass[aps,article,superscriptaddress,nofootinbib]{revtex4-1}
\usepackage{amsmath}
\usepackage{amsfonts}
\usepackage{dcolumn}
\usepackage{setspace}
\usepackage{color}
\usepackage{verbatim}
\usepackage{graphicx}

\definecolor{darkgreen}{rgb}{0,0.35,0}

\begin{document}

\title{Analytic self-gravitating Skyrmions, cosmological bounces and AdS
wormholes}
\author{Eloy Ay\'on-Beato}
\email{ayon-beato-at-fis.cinvestav.mx}
\author{Fabrizio Canfora}
\email{canfora-at-cecs.cl}
\author{Jorge Zanelli}
\email{z-at-cecs.cl}

\begin{abstract}
We present a self-gravitating, analytic and globally regular Skyrmion solution of the Einstein-Skyrme system with winding number $w=\pm1$, in presence of a cosmological constant. The static spacetime metric is the direct product $\mathbb{R} \times S^{3}$ and the Skyrmion is the self-gravitating generalization of the static hedgehog solution of Manton and Ruback with unit topological charge. This solution can be promoted to a dynamical one in which the spacetime is a cosmology of the Bianchi type-IX with time-dependent scale and squashing coefficients. Remarkably, the Skyrme equations are still identically satisfied for all values of these parameters. Thus, the complete set of field equations for the Einstein-Skyrme-$\Lambda$ system in the topological sector reduces to a pair of coupled, autonomous, nonlinear differential equations for the scale factor and a squashing coefficient. These equations admit analytic bouncing cosmological solutions in which the universe contracts to a minimum non-vanishing size, and then expands. A non-trivial byproduct of this solution is that a minor modification of the construction gives rise to a family of stationary, regular configurations in General Relativity with negative cosmological constant supported by an $SU(2)$ nonlinear sigma model. These solutions represent traversable AdS wormholes with NUT parameter in which the only ``exotic matter" required for their construction is a negative cosmological constant.
\end{abstract}

\maketitle

\affiliation{Departamento de F\'{\i}sica, CINVESTAV--IPN, Apdo. Postal 14--740, 07000 M\'exico~D.F., M\'exico} \affiliation{Centro de Estudios Cient\'{\i}ficos (CECS), Arturo Prat 514, Valdivia, Chile} \affiliation{Instituto de Ciencias F\'{\i}sicas y Matem\'aticas, Universidad Austral de Chile, Casilla 567 Valdivia, Chile}

\affiliation{Centro de Estudios Cient\'{\i}ficos (CECS), Arturo Prat 514, Valdivia, Chile}

\affiliation{Centro de Estudios Cient\'{\i}ficos (CECS), Arturo Prat 514, Valdivia, Chile}

\section{Introduction}
The Skyrme model \cite{skyrme} provides a useful description of nuclear and particle physics in the low energy regime of QCD \cite{witten0}. It allows the existence of a particular class of solitons ---\textit{Skyrmions}---, which represent Fermionic states in spite of the fact that the basic fields have spin zero. Skyrmions describe nucleons both theoretically and phenomenologically, where the identification of the winding number of the Skyrmion and the baryon number of particle physics is an essential observation \cite{witten0} (see, e.g. \cite{All,ANW,manton}). The possibility of treating the Skyrmions as fermions has been recently extended to curved spaces as well \cite{curved1f,curved2f}. Following \cite{lucock}, spherically symmetric black-hole solutions with a nontrivial Skyrme field --Skyrme ``hair''--- were found numerically in \cite{droz}. These are the first genuine counterexamples to the ``no-hair" conjecture; they are stable \cite{droz2} in contrast with other ``hairy" examples as the colored black holes (see e. g.,  \cite{Bizon:1994dh} for a review). Their regular particle-like counterparts  \cite{numerical1} and its dynamical properties have also been investigated numerically in \cite{numerical2}.

It would be very interesting and useful to go beyond numerical approximations in order to have explicit examples of analytic self-gravitating Skyrmions, that is, exact solutions on a spacetime that consistently solves the Einstein equations with the stress-energy tensor of the Skyrmion as the source. In particular, its time evolution would help to understand the gravitational consequences of having a discrete topological charge and the fact that these objects have a characteristic size. Unfortunately, there is no known analytic example of a four-dimensional self-gravitating configuration with non-vanishing topological charge. In fact, finding analytic spherically symmetric Skyrmion solutions even without considering their back-reaction to gravity is a highly non-trivial task \cite{manton}. The situation becomes even more difficult for the Einstein-Skyrme system and the search for analytic self-gravitating solutions with non-trivial topological charge may seem to be a problem beyond reach.

Here we show how to construct a self-gravitating Skyrmion in presence of a cosmological constant. This seems to be the first analytic solution of the Einstein-Skyrme system with nonvanishing winding number. In the static case, the spacetime metric is the Cartesian product $\mathbb{R}\times S^{3}$ for the time direction and the spatial slices, respectively.

The solution can be additionally promoted to a time-dependent one: a dynamical universe containing a self-gravitating Skyrmion. In this case there are bouncing solutions in which the universe contracts down to a nonvanishing minimum size, determined by the Skyrmion scale, and then expands again. A similar construction also gives rise to a family of exact stationary, regular configurations in General Relativity supported by the $SU(2)$ nonlinear sigma model with negative cosmological constant. These is a topologically non-trivial solution representing a traversable wormhole with NUT parameter. The only ``exotic matter" that violates the energy conditions needed to construct such wormholes is the negative cosmological constant.

\section{The self-gravitating Skyrmion}
We are interested in self-gravitating Skyrmions for the $SU(2)$ group described by the action {\small 
\begin{equation}
I[g,U]=\int d^{4}x\sqrt{-g}\left( \frac{R-2\Lambda}{2\kappa} +\frac{K}{4}  \mathrm{Tr}[A^\mu A_\mu +\frac{\lambda}{8}F_{\mu \nu }F^{\mu \nu}] \right). \label{skyrmaction}
\end{equation}
}Here $A_\mu$ is a shorthand for $U^{-1}\nabla_{\mu}U$, with $U\in SU(2)$ and $F_{\mu \nu}= \left[A_{\mu},A_{\nu}\right]$; $A_\mu =A_\mu^j t_j$ where $t_j=-i\sigma_j$ are the $SU(2)$ generators and $\sigma_{j}$ are the Pauli matrices. In our conventions $c=\hbar=1$, the spacetime signature is $(-,+,+,+)$ and Greek indices run over spacetime. Moreover, $R$ is the Ricci scalar, $\Lambda$ is the cosmological constant, $\kappa$ the gravitational constant, while phenomenology determines the positive Skyrme coupling constants $K$ and $\lambda$ \cite{ANW}.

The Skyrme and Einstein equations read 
\begin{subequations} \label{system}
\begin{align}
\nabla^{\mu}A_{\mu}+\frac{\lambda}{4}\nabla^{\mu}[A^{\nu},F_{\mu\nu}]&=0\ , \label{nonlinearsigma1} \\
G_{\mu \nu }+\Lambda g_{\mu \nu }& =\kappa T_{\mu \nu }\ ,  \label{einstein}
\end{align}
\end{subequations}
where $G_{\mu \nu }$ is the Einstein tensor and the Skyrme energy-momentum tensor is defined by 
\begin{equation}
T_{\mu\nu}= -\frac{K}{2}\mathrm{Tr} \left[ A_{\mu}A_{\nu }-\frac{1}{2} g_{\mu\nu}A^{\alpha}A_{\alpha}+ \frac{\lambda}{4}\left(g^{\alpha\beta}F_{\mu\alpha}F_{\nu\beta} -\frac{1}{4} g_{\mu\nu} F_{\alpha\beta} F^{\alpha\beta} \right)\right] .  \label{tmunu}
\end{equation}

We adopt the standard parametrization of the $SU(2)$-valued scalar $U(x^{\mu})$ \cite{Canfora}, as 
\begin{equation}
U^{\pm1}(x^\mu)=Y^0(x^\mu)\pmb{\mathbb{I}} \pm Y^i(x^\mu)t_i, \ \
\left(Y^0\right)^2 + Y^iY_i=1,  \label{standard1}
\end{equation}
where $\pmb{\mathbb{I}}$ is the $2\times2$ identity matrix. The unit vector $Y^{A}=(Y^{0},Y^{i})$ that defines the embedded three sphere, which is naturally given by 
\begin{subequations} \label{3-sphere}
\begin{align}  \label{pions1}
Y^0 &=\cos\alpha, & Y^i &=n^i \sin\alpha, &  \\
n^1 &=\sin\Theta \cos \Phi, & n^2 &=\sin \Theta \sin \Phi, & n^3 &=\cos\Theta.  \label{pions2}
\end{align}
\end{subequations}

The spacetime geometry for the static solutions of the coupled system (\ref{system}) is the product $\mathbb{R}\times S^{3}$, 
\begin{equation}
ds^{2}=-dt^{2}+\frac{\rho_{0}^{2}}{4}\left[ (d\gamma +\cos \theta d\varphi)^{2}+d\theta ^{2}+\sin ^{2}\theta d\varphi ^{2}\right] ,  \label{metric1}
\end{equation}
where $0\leq \gamma \leq 2\pi $, $0\leq \theta \leq \pi $, $0\leq \varphi \leq 2\pi $ are the coordinates on the 3-sphere and the constant $\rho _{0}$ is the radius. With this information one can solve (\ref{nonlinearsigma1}) for $\alpha $, $\Theta $ and $\Phi $ as functions of $\gamma $, $\theta $ and $\varphi $. It can be directly checked that the configuration 
\begin{equation}
\Phi =\frac{\gamma +\varphi }{2},\ \tan \Theta =\frac{\cot \left( \frac{\theta }{2}\right) }{\cos \left( \frac{\gamma -\varphi }{2}\right) },\ \tan \alpha =\frac{\sqrt{1+\tan ^{2}\Theta }}{\tan \left( \frac{\gamma -\varphi }{2}\right) },  \label{pions2.25}
\end{equation}
identically satisfies the Skyrme field equations (\ref{nonlinearsigma1}) on the background metric (\ref{metric1}). This was already noted long ago by Manton and Ruback \cite{curved} (see also \cite{bratek}). Those authors, however, did not attempt the construction of a consistent solution taking into account the back-reaction of the Skyrmion on the geometry. In order to do that, one should solve the Einstein equations (\ref{einstein}) with the stress-energy tensor (\ref{tmunu}) generated by the Skyrmion $U$ of the form (\ref{standard1}), (\ref{3-sphere}), (\ref{pions2.25}). The only nonvanishing components of $T_{\mu }^{\nu }$ are 
\begin{equation}
T_{t}^{t}=-\frac{3K(\lambda +\rho _{0}^{2})}{2\rho _{0}^{4}},\qquad T_{\gamma }^{\gamma }=T_{\theta }^{\theta }=T_{\varphi }^{\varphi }=\frac{K(\lambda -\rho _{0}^{2})}{2\rho _{0}^{4}}.  \label{tmunu0}
\end{equation}
It can be observed that although the solution $U$ explicitly depends on the angles $\gamma $, $\theta $ and $\varphi $, its energy-momentum tensor does not, which means that the Skyrmion respects spherical symmetry and therefore the back reaction need not upset the isometries of the background geometry (\ref{metric1}). Solving Einstein's equations with the energy-momentum tensor (\ref{tmunu0}) algebraically fixes the radius of the three-dimensional sphere and the Skyrme coupling constant in terms of the remaining coupling constants by the conditions 
\begin{equation}
\rho _{0}^{2}=\frac{3\left( 2-\kappa K\right) }{4\Lambda },\qquad \lambda =\frac{3\left( 2-\kappa K\right) ^{2}}{8\Lambda \kappa K},  \label{pions2.3}
\end{equation}
which requires $\Lambda (2-\kappa K)>0$. Hence, the metric (\ref{metric1}) together with the static Skyrmion (\ref{standard1}), (\ref{3-sphere}) and (\ref{pions2.25}) define a self-consistent solution of the full Einstein-Skyrme system (\ref{system}) provided the conditions (\ref{pions2.3}) are satisfied. This solution is the self-gravitating generalization of the Skyrmions in \cite{curved}. Our result can also be seen as a generalization of the hedgehog ansatz discussed in \cite{Canfora}, that allows for the construction of exact multi-Skyrmion configurations composed by elementary spherically symmetric Skyrmions with non-trivial winding number in four-dimensions \cite{Canforaetal}.

On any three-dimensional constant time hypersurface, the winding number for the configuration is 
\begin{equation}
w=\frac{-1}{24\pi^2} \int \mathrm{Tr}[\epsilon^{ijk}A_i A_j A_k ]=+1 \, , \label{winding}
\end{equation}
which implies that this Skyrmion cannot be continuously deformed to the trivial vacuum ($U=1$) \cite{manton}.

The above static Skyrmion can be promoted to a time-dependent solution in which the spacetime is a cosmology of the Bianchi type-IX , described by the metric 
\begin{equation}
ds^2={}-dt^2+\frac{\rho^2 (t)}{4}\left[ a^2 (t)(d\gamma +\cos \theta d\varphi )^{2}+d\theta ^{2}+\sin ^{2}\theta d\varphi ^{2}\right] , \label{dyn1}
\end{equation}
where $\rho (t)$ is a global scale factor and $a(t)$ is a squashing coefficient. The Skyrmion has the form (\ref{standard1}) with $Y^{0}$ and $Y^{i}$ also given by (\ref{3-sphere}). Remarkably, as one can check directly by hand, if the functions $\alpha $, $\Theta $ and $\Phi $ are again chosen as in the ansatz (\ref{pions2.25}), then the Skyrme field equations (\ref{nonlinearsigma1}) are still identically satisfied but now on the more general background (\ref{dyn1}). 

The technical reason why this happens is that the scale factor $\rho$ and the squashing parameter $a$ only depend on time, while the Skyrme ansatz depends on the spatial coordinates only. This surprising result is actually consistent with  an ansatz for the Skyrmion in which  the full Skyrme system is consistently reduced to a single scalar equation for the profile \cite{Canfora}. One can easily see that $\rho (t)$ and$a(t)$ respect the conditions in \cite{Canfora} to produce this ansatz. This is easier to see in the parametrization chosen in Eq. (\ref{pions2.25}) than in the standard parametrization used in \cite{curved}. 

Obviously, the Skyrme configuration in this case still has baryon charge $+1$. The energy momentum tensor is compatible with the Bianchi IX metric, as the only non-vanishing components are 
\begin{align}
T_{t}^{t}& =-\frac{K(a^{2}v+u)}{2\rho ^{4}a^{2}}, & T_{\varphi }^{\gamma }& =\frac{K(1-a^{2})(\lambda +\rho ^{2})\cos \theta }{\rho ^{4}a^{2}},  \notag \\
T_{\gamma }^{\gamma }& =-\frac{K(a^{2}v-u)}{2\rho ^{4}a^{2}}, & T_{\theta}^{\theta }& =T_{\varphi }^{\varphi }=\frac{K(\lambda a^{2}-\rho ^{2})}{2\rho ^{4}a^{2}},  \label{tmunue}
\end{align}
where $u=2\lambda +\rho ^{2}$ and $v=\lambda +2\rho ^{2}$. Direct computation reveals that the full Einstein-Skyrme system (\ref{system}) reduces then to the following three equations for $a$ and $\rho $ 
\begin{subequations}
\label{equd}
\begin{align}
2a\rho ^2 (2\rho \dot{a}+3a\dot{\rho})\dot{\rho}-2a^2 \rho^2 (\Lambda \rho^2 +a^2 -4)-\kappa K[(2\rho^2 +\lambda) a^2 +\rho^2 +2\lambda ]& =0,  \label{equd1} \\
2a^2 \rho^2 (2\rho \ddot{\rho}+\dot{\rho}^2)-2a^2 \rho^2 (\Lambda \rho^2 +3a^2 -4)-\kappa K[(2\rho^2 +\lambda)a^2 -\rho^2 -2\lambda]& =0,  \label{equd2} \\
a\rho^3 (\rho \ddot{a}+3\dot{\rho}\dot{a})+(a^2 -1)[\kappa K(\lambda +\rho^2)+4a^2\rho^2]& =0,  \label{equd3}
\end{align}
which correspond to the $t-t$, the $\gamma -\gamma $ and $\varphi -\gamma $ components of the Einstein equations. It can be easily checked that the remaining equations are not linearly independent and that Eq. (\ref{equd1}) is consistent with Eqs. (\ref{equd2}) and (\ref{equd3}). 

Although the full analysis of this system lies well beyond the scope of this paper, we will exhibit one consistent subfamily of cosmologies defined by fixing $a^{2}=1$, in which case (\ref{equd3}) is identically satisfied, while (\ref{equd1}) and (\ref{equd2}) reduce to 
\end{subequations}
\begin{align}
\dot{\rho}^2 & =\frac{\Lambda}{3}\rho^2 +\frac{\lambda \kappa K}{2\rho^2}+\frac{\kappa K-2}{2}\ ,  \label{firstint1} \\ \ddot{\rho}& =\frac{\Lambda }{3}\rho -\frac{\lambda \kappa K}{2\rho^3}\ .
\label{firstint2}
\end{align}
Eq.~(\ref{firstint1}) is a first integral of (\ref{firstint2}) and therefore the Einstein-Skyrme system reduces to an effective one-dimensional Newtonian problem for $\rho (t)$, with fixed \textquotedblleft total energy" $E=(\kappa K-2)/4$. It can be observed that in order to admit a static solution $\rho =\rho _{0}$, the parameters of the system must obey (\ref{pions2.3}). The qualitative behavior of the solution can be seen from the effective potential for the dynamical system, 
\begin{equation}
V=-\frac{\Lambda }{6}\rho ^{2}-\frac{\lambda \kappa K}{4\rho ^{2}}\,. \label{eff-pot}
\end{equation}
This potential is negative, unbounded below and approaches $-\infty $ both for $\rho^2 \rightarrow 0$ and $\rho^2\rightarrow \infty$. The maximum of $V$ corresponds to the saturating energy $E=(\kappa K-2)/4<0$. Below this energy, there are two possible scenarios for $\rho (t)$: an expansion from $\rho =0$ to a maximum $\rho =\rho _{max}$, and back to $\rho =0$. The second starts from $\rho =\infty $, bouncing at some minimal $\rho _{min}>0$ and then an accelerated expansion back to $\rho =\infty $. The turning points $                              \rho_{\ast }=\{\rho_{\min },\rho_{max}\}$ are determined by the parameters in the action, 
\begin{equation}
\rho_{\ast}^2=\frac{3}{4\Lambda }\left( 2-\kappa K\pm \sqrt{(2-\kappa K)^2 -\frac{8\lambda \Lambda K\kappa }{3}}\right) .
\end{equation}
These bouncing solutions occur within the range of parameters that ensures a reasonable cosmological evolution for the Skyrme model. The constraints were numerically determined in \cite{cosmo} for vanishing winding number $w$ where no bouncing was found, which suggests that the bouncing may be due to the presence of a soliton with nontrivial topological charge.

\section{AdS traversable wormholes}
In the special limit $\lambda =0$ the system reduces to the nonlinear sigma model. The nonlinear sigma model is an important non-linear field theory with applications to a wide range of phenomena, from quantum field theory to statistical mechanics systems like quantum magnetism, the quantum hall effect, super fluid $^{3}$He, and string theory \cite{manton}. The most relevant application in particle physics is the $SU(2)$ non-linear sigma model for the description of the low-energy dynamics of Pions in 3+1 dimensions (see for a detailed review \cite{Nair}).

Redefining $\gamma $ as a time coordinate $\tau $ and $t$ as a spatial coordinate $z$, the previous configuration gives rise to new stationary solutions 
\begin{align}
ds^{2}= \frac{\rho^2}{4}(z)\left[ -Q^{2}(d\tau +\cos \theta d\varphi)^{2} + (d\theta ^{2}+\sin ^{2}\theta d\varphi ^{2})\right] +dz^{2}, \label{metricnew}
\end{align}
with NUT parameter $Q$. The solution is single-valued provided the NUT parameter is an even integer, $|Q|=2N$, and $0\leq \tau \leq 2\pi N$ \cite{griffpod}. It can be directly checked that in any background of the form (\ref{metricnew}), the configuration defined in (\ref{3-sphere}) and (\ref{pions2.25}) with $\gamma $ substituted by $\tau $ identically satisfies the field equations for the non-linear sigma model. The corresponding energy-momentum tensor, 
\begin{align}
T_\tau ^\tau & =-\frac{K(2Q^2 +1)}{2Q^2 \rho^2}, & T_z ^z & =- \frac{K\left( 2Q^2 -1\right)}{2Q^2\rho^2},  \notag \\
T_\theta ^\theta & =T_\varphi ^ \varphi =\frac{K}{2Q^2 \rho^2}, &  T_\varphi ^\tau & =-\frac{K\left( Q^2 +1\right) }{Q^ 2\rho^2}\cos\theta ,  \label{tmunu1.1}
\end{align}
has positive energy density. In fact, the full $T_{\mu }^{\nu }$ of the nonlinear sigma model satisfies both the null and the weak energy conditions  \cite{energycond}. It should be noted that the present configuration of the non-linear sigma model can be seen as a topologically non-trivial generalization of the so-called \textit{boson stars ansatz} for complex scalar fields \cite{Herdeiro}, which are time-dependent and yet their stress-energy tensor is time-independent so that they can naturally live in an stationary spacetime background. The configuration defined in (\ref{3-sphere}) and (\ref{pions2.25}) with $\gamma $ substituted by $\tau $ depends explicitly on $\tau $ and, nevertheless, the corresponding energy-momentum tensor (\ref{tmunu1.1}) is compatible with a stationary metric. It can be directly checked that the Einstein equations are satisfied provided $\Lambda <0$, 
\begin{equation}
\rho (z)=\sqrt{\frac{3(\kappa K-8)}{4|\Lambda |}}\cosh\left( \frac{|\Lambda|^{1/2}}{\sqrt{3}} z\right) \ ,  \label{rho}
\end{equation}
and $Q^{2}=\frac{\kappa K}{4}$. This solution corresponds to a sector in the parameter space of the action (\ref{skyrmaction}) characterized by $\lambda=0$, $\kappa K>8$, $\Lambda <0$, which is not overlapping with the sector of the theory that admits the Skyrmion solutions ($\lambda >0$, $\kappa K<2$, $\Lambda >0$).

The stationary geometries (\ref{metricnew}) describe wormholes with NUT parameter in General Relativity with negative cosmological constant and the nonlinear sigma model as source. These wormholes have asymptotic NUT-AdS regions for $z\rightarrow \pm \infty $ connected by a throat at $z=0$. The geometry is locally regular and has no curvature singularities but, as it is well known, in the presence of a NUT parameter the causal structure includes closed time-like curves \cite{griffpod} It has been recently shown, however, that space-times with NUT parameter can be defined as geodesically complete even without periodic time and no closed causal geodesics \cite{clementNUT}. This means that it might be possible to avoid some longstanding obstructions to accept the Taub-NUT solution as physically relevant.

These wormholes are also traversable in the sense that there exist timelike geodesics that connect both sides of the throat \cite{visser-lobo}. This is confirmed by analyzing, for example, radial geodesics. Indeed, consider the radial a time-like geodesic (parametrized by $s$) whose effective Lagrangian is 
\begin{equation*}
L_{\mbox{eff}}=g_{\mu \nu }(q) \dot{q}^\mu \dot{q}^\nu ,\ q^\mu =(\tau(s) , z(s) , 0 , 0 ),\ \dot{q}^\mu = \frac{d}{ds}q^\mu .
\end{equation*}
The system has two integrals of motion: the Lagrangian itself and the Noether charge associated to the Killing vector $\partial_\tau $, 
\begin{equation*}
g_{\mu \nu} \dot{q}^\mu \dot{q}^\nu = - I^2\ ,\ \ \partial _{s}\left[ \frac{\partial L_{\mbox{eff}}}{\partial \dot{\tau}}\right] =0\ \Rightarrow \ \frac{\partial L_{\mbox{eff}}}{\partial \dot{\tau}}=E \ . 
\end{equation*}
With the above integrals of motion and after simple manipulations, the radial ($z$) geodesic becomes 
\begin{equation*}
\left( \frac{dz}{ds}\right)^2 -\frac{E^2 }{Q^2 \rho ^2 (z)}=-I^2 \ , 
\end{equation*}
where $\rho$ is given in (\ref{rho}), can be analytically integrated. Hence, the radial geodesics describe a bounded periodic motion around $z=0$ within an attractive effective potential. For small $I^{2}$ the geodesics enter deeply into the positive-$z$ and negative-$z$ regions. Consequently, the above wormholes are traversable. The fact that the orbits are bounded is a consequence of the negative cosmological constant, the only ``exotic matter" needed to support these wormholes.

According to standard no-go theorems there can be no stationary traversable wormholes in four-dimensional general relativity minimally coupled with physical non-exotic sources \cite{visser-lobo,Butcher}. Hence, the existence of wormholes requires either the well known violations of the energy conditions, or that the source itself should be non stationary. The boson star-like ansatz for the non-linear sigma model presented here, is $\tau$-dependent and therefore fulfills the requirement, while the corresponding energy momentum tensor is still compatible with a stationary metric.

\section{Conclusions and perspectives}
The solution presented above seems to be the first self-gravitating Skyrmion, an analytic, globally regular solution of the four-dimensional Einstein-Skyrme with winding number $1$. The static space-time metric is the Cartesian product of $\mathbb{R}$ (for the time-like direction) times $S^{3}$ for the space-like slices. The Skyrmion is constructed using the generalized hedgehog ansatz with a non-trivial profile. This solution can be further promoted to a dynamical Bianchi IX cosmological solution with two independent scale factors while the Skyrme field equations are self-consistently satisfied using the generalized hedgehog ansatz with winding number $1$. In this sector, the complete field equations of the four-dimensional Einstein-Skyrme system consistently reduces to a pair of autonomous differential equations for the two scale factors which encodes the presence of a conserved topological charge. The resulting cosmologies include \textit{bouncing solutions} in which the scale factor contracts to a minimal non-vanishing size and then expands again. Other solutions are Big-Bang cosmologies with a maximal expansion and a final big crunch.

Using a similar technique, an exact static traversable Lorentzian wormhole in four-dimensional Einstein theory, minimally coupled to the nonlinear sigma model with a negative cosmological constant was found. There is no "exotic matter" (as a negative cosmological constant can hardly be considered exotic) and yet this wormhole is traversable, Lorentzian and built purely with ingredients arising in standard particle physics and with $\Lambda<0$. The ansatz for the nonlinear sigma model is everywhere regular, depends explicitly on time, circumventing Derrick's theorem; however, the dependence is such that its energy-momentum tensor is compatible with a stationary metric.

The present work can be seen as an investigation in the framework of the so-called scalar-tensor theories \cite{ST}, in which one or more scalar fields are included in the action to account for some modifications to pure gravity. This may be useful in cosmological scenarios \cite{cosmo1}, or to describe the dynamics of dense astrophysical objects such as neutron stars or, more generally hadronic matter stars, including supernovae near the collapse into black holes \cite{hadronic}. The standard approach is to start from a given spacetime background and try to find solutions for the matter field in it, possibly including some adjustments.

Our strategy here is the converse: we start with a topologically stable highly symmetric matter configuration, a solitonic state of the Skyrme theory or sigma model, and then look for the geometry which can support it. In fact, one could have applied this approach to find the Reisner-Nordstrom (RN) solution by first identifying the static, spherically symmetric solution of Maxwell's equations ($A=qr^{-1}dt$) and then find the geometry where this solution can be accommodated; the answer is, of course, the RN black hole. In the present case, one can appreciate that the matter configuration is nontrivial because of the underlying $SU(2)$ symmetry, which in turn requires three scalar fields in order to have a nontrivial mapping into the spatial section of spacetime, so that one or two scalar fields would have been insufficient. Another peculiarity of the solution is that the Skyrmion is not a perturbative probe covering the spacetime that could be switched off. This is highlighted by the fact that the Skyrmion has nonvanishing winding number.

Interesting issues worth further investigation, including the role of the topological charge in the bouncing cosmologies, the stability of the full dynamical system (\ref{equd}), and the global causal structure of the wormhole solutions, will be discussed in a forthcoming work \cite{EFJ-2}.

\begin{acknowledgments}
This work has been funded by the Fondecyt grants 1120352, 1121031, 1130423, 1140155 and 1141073, together with the CONACyT grants 175993 and 178346. EAB was partially supported by the ``Programa Atracci\'{o}n de Capital Humano Avanzado del Extranjero, MEC'' from Conicyt. The Centro de Estudios Cient\'{\i}ficos (CECs) is funded by the Chilean Government through the Centers of Excellence Base Financing Program of Conicyt.
\end{acknowledgments}

\end{document}